# Quantum nonlinear cavity QED with coherently prepared atoms

Guoqing Yang[1,4], Wen-ju Gu [2-3], Gaoxiang Li[2], Bichen Zou[1], and Yifu Zhu[1]

[1]Department of Physics, Florida International University, Miami, Florida 33199
[2]College of Physics, Huazhong Normal University, Wuhan, 430079, China
[3]School of Physics and Optoelectronic engineering, Yangtze University, Jinzhou, 434023, China
[4] School of Electronic and Information, Hangzhou Dianzi University, Hangzhou 310018,China

Abstract

We propose a method to study the quantum nonlinearity and observe the multiphoton transitions in a multiatom CQED system. We show that by inducing simultaneously destructive quantum interference for the single-photon and two-photon excitations in the CQED system, it is possible to observe the direct three-photon excitation of the higher-order ladder states of the CQED system. We report an experiment with cold Rb atoms confined in an optical cavity and demonstrate such interference control of the multi-photon excitations of the CQED system. The observed nonlinear excitation of the CQED ladder states agrees with a theoretical analysis based on a fully quantized treatment of the CQED system, but disagrees with the semiclassical analysis of the CQED system. Thus it represents the first direct observation of the quantum nature of the multiatom CQED system and opens new ways to explore quantum nonlinearity and its applications in quantum optical systems in which multiple absorbers/emitters are coupled with photons in confined cavity structures.

I. Introduction

Cavity quantum electrodynamics (CQED) studies the fundamental atom-photon interactions and has important applications in quantum physics and quantum electronics [1]. A variety of the CQED systems has been realized [2-5], and a wide array of fundamental studies and practical applications based on the CQED concepts and effects have been explored [6-11]. A basic CQED system consists of single atoms/atomic qubits coupled to a cavity mode. Studies of CQED can also be done with a composite system consisting of a cavity mode collectively coupled with multiple atoms/atomic qubits. The CQED system with multiple atoms is uniquely suited for studies of collective atom and photon interactions, in which the Dicke states of the atoms and the cavity mode form the collective polariton states and lead to interesting physical phenomena such as quantum many body effects [6], quantum entanglement of multiple atoms [7], and a cavity controlled supperradiant laser [11].

It was recognized that the linear excitation of the CQED system reveals two normal modes with a frequency separation commonly referred to as the vacuum Rabi splitting and can be understood classically as two coupled linear oscillators [12]. To reveal the intrinsic quantum mechanical nature of the CQED system and explore potential applications of the CQED quantum nonlinearities [13-15], it is necessary to induce the multiphoton transitions in the higher-order ladder states of the CQED system and observe the quantum nonlinearity in CQED system. In recent years, the multiphoton transitions associated with the quantum nonlinearity have been observed in the single-atom/qubit CQED systems [16-19]. In a multiatom CQED system, the collective polariton states of the CQED system form a ladder system with equal spacing among different orders and the multiphoton transitions [20] become degenerate in the transition frequency with the single photon transition [21]. Although the resonant multiphoton excitation is now possible, it is difficult to separate the dominant single photon transition from the multiphoton transitions, and explore the quantum nonlinearity and its applications in the multiatom CQED system. Although the earliest observation of the vacuum Rabi splitting was reported in a CQED system with multiple atoms decades ago [1] and there are theoretical proposals to study the quantum nonlinear excitation in CQED systems with a few atoms [22-23], it is still elusive to attempt the experimental observation of the direct multiphoton transitions in a multiatom CQED system.

Here we propose a method to study the quantum nonlinear CQED in a coupled cavity and multiatom system and observe the pure three-photon transition of the quantum ladder states of the CQED system. The method replies on inducing simultaneously the quantum destructive interference for the single-photon transition and two-photon transition in the multi-ladder CQED system which suppresses both the single-photon and two-photon transitions and resonantly enhances the three-photon transition. We present experimental results that demonstrate such interference technique for studies of the quantum nonlinearities in the multiatom CQED system. The experimental measurements agree with the theoretical analysis based on a fully quantized treatment of the CQED system, but disagree with the semiclassical analysis of the cavity QED system. Thus our experimental work represents the first observation of the pure three-photon transition in the quantum ladder states and direct demonstration of the quantum nature of the multi-atom CQED system.

II. Theoretical analysis

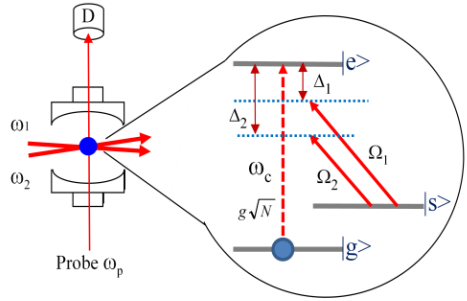

Fig.1 Coherently coupled multi-atom CQED system in which N three-level atoms are coupled to the cavity mode and two free-space laser fields $\omega_1$ and $\omega_2$. $\Omega_1$ and $\Omega_2 \ll g\sqrt{N}$.

Fig. 1 shows the schematic diagram for the multi-atom CQED system that consists of a single mode cavity containing N three-level atoms interacting with two coupling lasers from free space. The cavity mode couples the atomic transition |g>-|e> and the classical coupling lasers drive the atomic transition |s>-|e> with Rabi frequency $2\Omega_1$ and $2\Omega_2$, respectively. A weak probe laser $\omega_p$ is coupled into the cavity and the transmitted probe light through the cavity versus its frequency reveals the excitation spectrum of the CQED system. We treat the input probe field and the cavity field quantum mechanically, but the coupling fields classically. The interaction Hamiltonian can be written as

$$H = \omega_e \sum_j \sigma_{ee}^{(j)} + \omega_g \sum_j \sigma_{gg}^{(j)} + \omega_s \sum_j \sigma_{ss}^{(j)} + \omega_c a^+ a + g \sum_j (a\sigma_{eg}^{(j)} + a^+ \sigma_{ge}^{(j)}) + \sum_j [(\Omega_1 e^{-i\omega_1 t} + \Omega_2 e^{-i\omega_2 t})\sigma_{es}^{(j)} + H.C.] + \alpha_p (a e^{i\omega_p t} + a^+ e^{-i\omega_p t}). \quad (1)$$

The last term in Eq. (1) represents the coupling of a weak probe field and the cavity mode. Here $\hat{a}$ is the annihilation operator of the cavity photons, $\sigma_{mn}^{(j)} = |m^{(j)}\rangle\langle n^{(j)}|$, (m,n=e, g, and s) are the atomic operators for the jth atom, $g = \mu_{eg}\sqrt{\omega_c / 2\hbar\varepsilon_0 V}$ is the cavity-atom coupling coefficient, and $\alpha_p$ is the probe field amplitude. We consider the cavity frequency matches the atomic transition frequency, i.e., $\omega_c = \omega_e - \omega_g = \omega_{eg}$, and define the coupling frequency detunings $\Delta_1 = \omega_1 - \omega_{es}$ and $\Delta_2 = \omega_2 - \omega_{es}$ ($\omega_{es} = \omega_e - \omega_s$). The collective atomic operators can be defined as $J_+ = \sum_j \sigma_{eg}^{(j)}$ and $R_+ = \sum_j \sigma_{es}^{(j)}$ (assuming the uniform coupling strength for the N identical atoms inside the cavity). Then in the interaction picture, $H_I^{(1)} = g(J_+ a + a^+ J_-) + [(\Omega_1 e^{-i\Delta_1 t} + \Omega_2 e^{-i\Delta_2 t})R_+ + H.C.]$ and $H_I^{(2)} = \alpha_p (a e^{i\Delta_p t} + a^+ e^{-i\Delta_p t})$ where $\Delta_p = \omega_p - \omega_c$ is the probe frequency detuning. For a weak probe field, the atomic population is concentrated in the ground state |g> and the collective atomic operators can be written in terms of the Dicke states $|\frac{N}{2}, -\frac{N}{2} + l\rangle$ [24] as

$$J_+ = \sum_{l,n} |\frac{N}{2}, -\frac{N}{2} + l\rangle\langle\frac{N}{2}, -\frac{N}{2} + l | J_+ | \frac{N}{2}, -\frac{N}{2} + n\rangle\langle\frac{N}{2}, -\frac{N}{2} + n|$$

$$= \sum_{l,n} |\frac{N}{2}, -\frac{N}{2} + l\rangle\langle\frac{N}{2}, -\frac{N}{2} + n|\sqrt{(N-n)(n+1)}\delta_{l,n+1} = \sum_n |\frac{N}{2}, -\frac{N}{2} + n+1\rangle\langle\frac{N}{2}, -\frac{N}{2} + n|\sqrt{(N-n)(n+1)}. \quad (2)$$

Here $l$ and $n$ are the atomic excitation numbers. Denote $|\frac{N}{2}, -\frac{N}{2} + n\rangle$ as $|-\frac{N}{2} + n\rangle$, then $J_+ = \sum_n |-\frac{N}{2} + n+1\rangle\langle -\frac{N}{2} + n|\sqrt{(N-n)(n+1)}$. Under the condition of $\Omega_1$ and $\Omega_2 \gg g$ (the atomic population in the state |s> is very small), we can define the collective atomic state $|s\rangle = \frac{1}{\sqrt{N}} \sum_j |g^{(1)}, \ldots, s^{(j)}, \ldots g^{(N)}\rangle$, and obtain

$$R_+ |s\rangle = \sum_{jj'} |e^{(j)}\rangle\langle s^{(j')}| \frac{1}{\sqrt{N}} |g^{(1)}, \ldots, s^{(j)}, \ldots g^{(N)}\rangle = \frac{1}{\sqrt{N}} \sum_j |g^{(1)}, \ldots, e^{(j)}, \ldots g^{(N)}\rangle = |-\frac{N}{2} + 1\rangle. \quad (3)$$

Then one derives

$$R_+ = |-\frac{N}{2} + 1\rangle\langle s| = \sum_m |-\frac{N}{2} + 1, m\rangle\langle s, m|, \quad (4)$$

and the annihilation operator of the cavity photons can be written in the Fock state basis as $a = \sum_m |m\rangle\langle m+1|\sqrt{m+1}$ (m is the Fock state number):

$$a = \sum_{nm}\sqrt{m+1}\,|m\rangle\langle m+1| = \sum_{mn}\sqrt{m+1}(|-\frac{N}{2}+n,m\rangle\langle-\frac{N}{2}+n,m+1|+|s,m\rangle\langle s,m+1|), \quad (5)$$

When the cavity is resonant with the atomic transition, $\omega_c = \omega_e - \omega_g = \omega_{eg}$, the system Hamiltonian can then be written as

$$H_I = g\sum_{nm}\sqrt{(N-n)(n+1)(m+1)}[|-\frac{N}{2}+n+1,m\rangle\langle-\frac{N}{2}+n,m+1|+H.C.]+\sum_m[(\Omega_1 e^{-i\Delta_1 t}+\Omega_2 e^{-i\Delta_2 t})|-\frac{N}{2}+1,m\rangle\langle s,m|+H.C.]$$

$$+\sum_{nm}\alpha_p\sqrt{m+1}[(|-\frac{N}{2}+n,m\rangle\langle-\frac{N}{2}+n,m+1|+|s,m\rangle\langle s,m+1|)e^{i\Delta_p t}+H.C.]. \quad (6)$$

The CQED system satisfies $g^2 N > \kappa\Gamma$ ($\kappa$ is the cavity decay rate and $\Gamma$ is the decay rate of the excited state $|e\rangle$) and is in the strong coupling regime for the collectively coupled atoms and the cavity mode (but $g^2 \ll \kappa\Gamma$; the CQED system is in the weak coupling regime for single atom and the cavity mode). The cavity photon and the atoms form the symmetric, Dicke-type atomic and photonic product states. The ground state of the cavity-atom system is $|-\frac{N}{2},0\rangle$ (all atoms are in the ground state and no photon in the cavity mode), the two product states with one excitation quanta are $|-\frac{N}{2},1\rangle$ (one photon in the cavity mode and all atoms are in the ground state $|g\rangle$) and $|-\frac{N}{2}+1,0\rangle$ (one atom in the excited state $|e\rangle$ and zero photon in the cavity mode); there are three product states with two excitation quanta, and four product states with three excitation quanta, and etc as shown in Fig. 2(a).

The interaction term of the cavity photons and the collective atomic operators is then given by $H_{int} = g(aJ_+ + a^+J_-) = g\sum_{nm}\sqrt{(N-n)(n+1)(m+1)}(|-\frac{N}{2}+n+1,m\rangle\langle-\frac{N}{2}+n,m+1|+H.C.)$. When the two free-space coupling fields drive the transition $|s\rangle - |e\rangle$ satisfy the condition $\Omega_1$ and $\Omega_2 \ll \sqrt{N}g$, they can be treated as perturbations to the cavity coupled atomic system. Then we can diagonalize the interaction Hamiltonian $H_{int} = g(aJ_+ + a^+J_-) = g\sum_{nm}\sqrt{(N-n)(n+1)(m+1)}[|-\frac{N}{2}+n+1,m\rangle\langle-\frac{N}{2}+n,m+1|+H.C.]$ and derive the eigenvalues and eigenstates of the CQED system. The eigenstates are the superposition of the atomic Dicke states and the cavity photon states, and are referred to as polariton states, which form an infinite ladder starting from the ground state J=0 (J=0, 1,… n.., is the excitation quanta number). For a given excitation J, there are J+1 polariton states with the energies given by $E_m = J\hbar\omega_c + m\sqrt{N}g$ (m=0, ±2, …±J for J is even, or m==±1, …±J for J is odd) as shown in Fig. 2(b).

The two free-space coupling lasers (with $\Delta_1 = -g\sqrt{N}$ and $\Delta_2 = -2g\sqrt{N}$) drive the same transition $|s\rangle - |e\rangle$ and create an infinite ladder of Floquet states $|s,m\rangle = \sum_j C_{s,n} |s^{(j)},m\rangle e^{ij\sqrt{N}gt}$ (with the equal frequency separation $\Delta_1 - \Delta_2 = \sqrt{N}g$ between the neighboring Floquet states) [25]. Here m is the number of the intra-cavity photons and j is the index for the jth-order Floquet state. Because $\sqrt{N}g \gg \Omega_1$ and $\Omega_2$, only two Floquet states $|s^{(0)},m\rangle$ and $|s^{(1)},m\rangle e^{i\sqrt{N}gt}$ (with j=0 and j=1) need to be considered and the rest of them can be neglected. This results in $|s,m\rangle \approx |s^{(0)},m\rangle + \frac{(\Omega_1 + \Omega_2)^2}{g^2 N} |s^{(1)},m\rangle e^{i\sqrt{N}gt}$. In the weak excitation regime (the probe laser coupled into

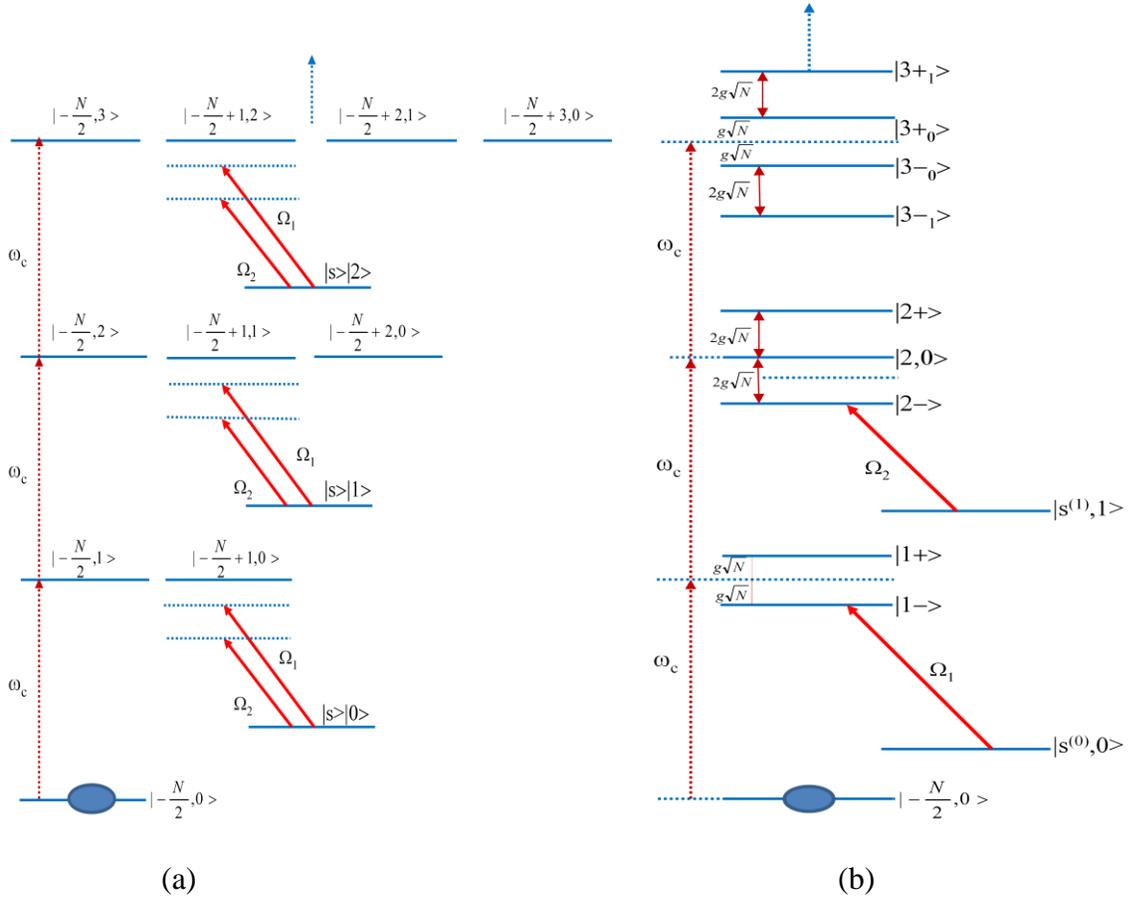

(a) (b)

Fig. 2  (a) The energy level diagram of the CQED product states. (b) The eigenstates ladder of the CQED system and the two free-space coupling fields connecting the atomic state |s> with the specific ladder states and the coupling laser detunings are $\Delta_1 = -g\sqrt{N}$ and $\Delta_2 = -2g\sqrt{N}$.

the cavity mode is very weak), we keep only up to J=3 excitation processes of the CQED system and neglect the other higher-order (J>3) processes. Then the CQED system can be approximately treated with the truncated basis consisting of 12 states as shown in Fig. 2(b), which are: the ground (J=0) state $|-\frac{N}{2},0\rangle = |0,0\rangle$; three first-order (J=1) excited polariton states, $|1_+\rangle = \frac{1}{\sqrt{2}}(|-\frac{N}{2}+1,0\rangle + |-\frac{N}{2},1\rangle$, $|1_-\rangle = \frac{1}{\sqrt{2}}(|-\frac{N}{2}+1,0\rangle - |-\frac{N}{2},1\rangle$, and $|s^{(0)},0\rangle$; four 2$^{nd}$-order (J=2) excited polariton states, $|2_+\rangle = \frac{1}{2}(|-\frac{N}{2}+2,0\rangle + \sqrt{2}|-\frac{N}{2}+1,1\rangle + |-\frac{N}{2},2\rangle)$, $|2_0\rangle = \frac{1}{\sqrt{2}}(|-\frac{N}{2}+2,0\rangle + |-\frac{N}{2},2\rangle)$, and $|2_-\rangle = \frac{1}{2}(|-\frac{N}{2}+2,0\rangle - \sqrt{2}|-\frac{N}{2}+1,1\rangle + |-\frac{N}{2},2\rangle)$, and $|s^{(1)},1\rangle$; and four 3$^{rd}$-order (J=3) excited polariton states, $|3_{-1}\rangle = \frac{1}{2\sqrt{2}}(-|-\frac{N}{2}+3,0\rangle + \sqrt{3}|-\frac{N}{2}+2,1\rangle - \sqrt{3}|-\frac{N}{2}+1,2\rangle + |-\frac{N}{2},3\rangle)$, $|3_{-0}\rangle = \frac{1}{2\sqrt{2}}(\sqrt{3}|-\frac{N}{2}+3,0\rangle - |-\frac{N}{2}+2,1\rangle - |-\frac{N}{2}+1,2\rangle + \sqrt{3}|-\frac{N}{2},3\rangle)$, $|3_{+0}\rangle = \frac{1}{2\sqrt{2}}(-\sqrt{3}|-\frac{N}{2}+3,0\rangle - |-\frac{N}{2}+2,1\rangle + |-\frac{N}{2}+1,2\rangle + \sqrt{3}|-\frac{N}{2},3\rangle)$, $|3_{+1}\rangle = \frac{1}{2\sqrt{2}}(|-\frac{N}{2}+3,0\rangle + \sqrt{3}|-\frac{N}{2}+2,1\rangle + \sqrt{3}|-\frac{N}{2}+1,2\rangle + |-\frac{N}{2},3\rangle)$. With the truncated 12 basis states, the system Hamiltonian is reduced to

$$H_{eff} = -g\sqrt{N}(|1_-\rangle\times\langle 1_-| + |s^{(0)},0\rangle\times\langle s^{(0)},0|) - 2g\sqrt{N}(|2_-\rangle\times\langle 2_-| + |s^{(1)},1\rangle\times\langle s^{(1)},1|) - 3g\sqrt{N}(|3_{-}\rangle\times\langle 3_{-}| - \Omega_1/2(|1_-\rangle\times\langle s^{(0)},0| + |s^{(0)},0\rangle\times\langle 1_-|) - \Omega_2/2(|2_-\rangle\times\langle s^{(1)},1| + |s^{(1)},1\rangle\times\langle 2_-|)$$

$$\alpha e^{i\Delta_p t}(\frac{1}{\sqrt{2}}|-\frac{N}{2},0\rangle\times\langle 1_-| + \frac{1}{\sqrt{2}}|1_-\rangle\times\langle 2_-| + \frac{\sqrt{6}}{2}|2_-\rangle\times\langle 3_-| + \frac{(\Omega_1+\Omega_2)^2}{g^2 N}||s^{(0)},0\rangle\times\langle s^{(1)},1|) + H.C. \quad (7)$$

Then one can derive the density matrix equations $\frac{d\rho}{dt} = i[\rho, H_{eff}] + L\rho$ where L is the damping operator from the atomic decay and cavity decay. The density matrix equations can then be solved numerically with the Quantum Optics Toolbox [26] and the expectation value of the intra-cavity photon number is given by (see the supplement material for details)

$$\langle a^+ a \rangle = \frac{1}{2}\rho_{1-,1-} + \rho_{s1,s1} + \rho_{2-,2-} + \frac{2}{3}\rho_{3-,3-}. \quad (8)$$

In Eq.(8), the first two terms, $\frac{1}{2}\rho_{1-,1-}$ and $\rho_{s1,s1}$, represent the contribution from the single-photon process; the 3$^{rd}$ term, $\rho_{2-,2-}$, represents the contribution from the two-photon process; and the 4th term, $\frac{2}{3}\rho_{3-,3-}$, represents the contribution from the three-photon process. The total number of photons transmitted through the cavity is then given by $\kappa\langle a^+ a\rangle$.

The excitation spectrum of the CQED system can be measured by coupling a weak probe laser into the cavity mode and collecting the transmitted probe photons while scanning its frequency detuning $\Delta_p = \omega_p - \omega_{ge}$. Without the coupling fields, the energy ladder of the CQED system is shown in Fig. 3(a). The spectrum (see Fig. 5(a)) exhibits two peaks located at $\Delta_p = \pm g\sqrt{N}$ where all

orders of multiphoton transitions are degenerate. However, with a weak probe laser far below the

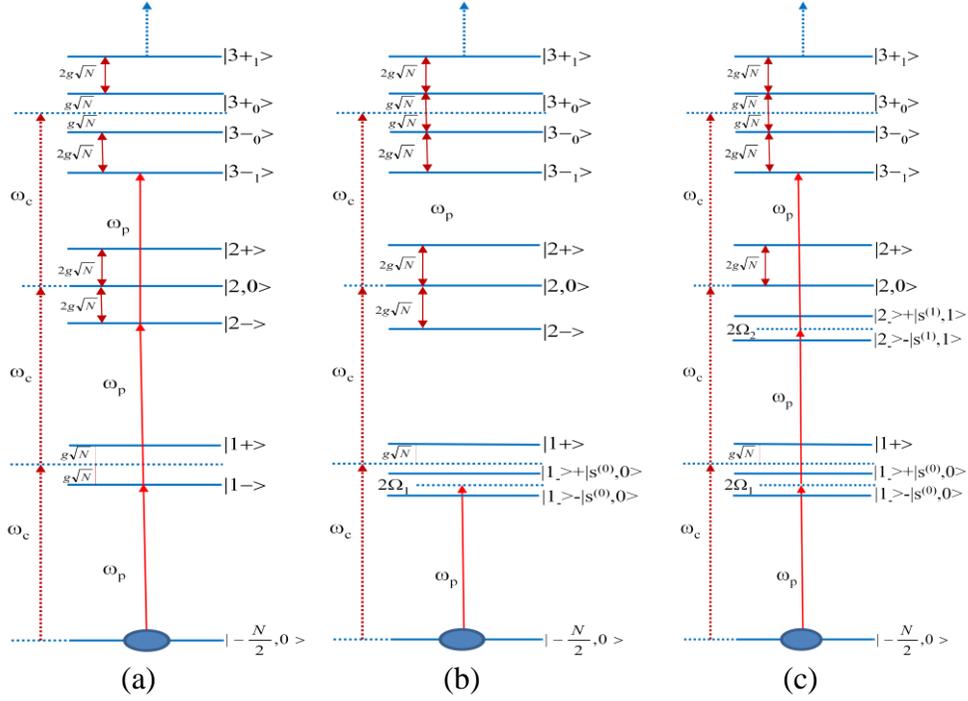

Fig. 3 (a) Without the free-space coupling fields, the quantum ladder states of the CQED system representing the multiple two-level atoms coupled to a single cavity mode. (b) With only a single coupling field ($\Omega_1 \neq 0$ but $\Omega_2 = 0$) tuned to the resonance of the transition $|s,0\rangle$ - $|1-\rangle$ ( $\Delta_1 = -g\sqrt{N}$ ). The coupling field produces two dressed states $|\Psi_\pm^1\rangle = \frac{1}{\sqrt{2}}(|1_-\rangle \pm |s^{(0)},0\rangle)$ separated by the Rabi frequency $2\Omega_1$, which leads to two excitation paths for the probe laser. The destructive interference between the two transition paths suppresses all orders of the transitions at $\Delta_p = -g\sqrt{N}$. (c) With $\Omega_1$ and $\Omega_2 \ll g\sqrt{N}$, the two coupling fields creates the dressed states $|\Psi_\pm^1\rangle = \frac{1}{\sqrt{2}}(|1_-\rangle \pm |s^{(0)},0\rangle)$ and $|\Psi_\pm^2\rangle = \frac{1}{\sqrt{2}}(|2_-\rangle \pm |s^{(1)},1\rangle)$, which opens the two excitation paths for the single photon transition and the two photon transition for a weak probe laser coupled into the cavity mode. When the probe laser is tuned to $\Delta_p = -g\sqrt{N}$, the single photon and two photon transitions are suppressed but the three photon transition, $|-\frac{N}{2},0\rangle$ -$|2_-\rangle$-$|3_-\rangle$, is then resonantly enhanced.

saturation, the single-photon transition is dominant and the two spectral peaks $\Delta_p = \pm g\sqrt{N}$ represent the resonant single-photon excitation of the first-order excited states (the polariton states or the normal modes) $|1_\pm\rangle$ [27-28].

When there is only one coupling laser present ($\Omega_1 \neq 0$ but $\Omega_2 = 0$) and it is tuned to the polariton resonance at $\Delta_1=-g\sqrt{N}$ (or $\Delta_1=+g\sqrt{N}$), the coupling laser 1 creates two dressed polariton states $|\Psi_\pm\rangle=\frac{1}{\sqrt{2}}(|1_\pm\rangle\pm|s,0\rangle)$ (Since $\Omega_1\ll\sqrt{N}g$, the effect of the coupling field 1 on other ladder states of the CQED system can be neglected due to the large detunings from these states) (see Fig. 3(b)), which results in two excitation paths with a π phase shift. The destructive interference suppresses all orders of the linear and nonlinear excitations (this configuration is similar to the EIT suppression of both single-photon and two-photon absorptions in a ladder type four-level atomic system in free space [29-31]) and a narrow dip appears in the spectral peak at $\Delta_p=-g\sqrt{N}$ as shown in Fig. 5(b) and reported in ref. [32].

When both coupling lasers are present ($\Omega_1\neq0$ and $\Omega_2\neq0$), and the coupling 1 is tuned to the polariton resonance $|1_-\rangle$ at $\Delta_1=-g\sqrt{N}$ (or $\Delta_1=+g\sqrt{N}$) and the coupling 2 is tuned to the next higher-order resonance $|2_-\rangle$ at $\Delta_2=-2g\sqrt{N}$ (or $\Delta_1=+2g\sqrt{N}$), (again, since both $\Omega_1$ and $\Omega_2\ll\sqrt{N}g$, the effects of the two coupling fields on other ladder states can be neglected due to the large detunings from these states). The coupling laser 1 creates two dressed polariton states

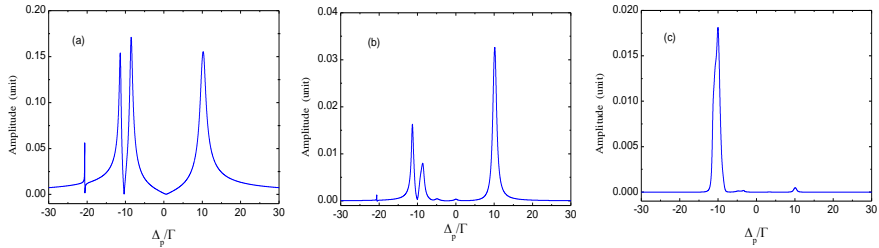

Fig. 4 With two coupling fields present ($\Omega_1=\Omega_2=\Gamma$, and $\Delta_1=-g\sqrt{N}$ and $\Delta_2=-2g\sqrt{N}$) are present, (a) the calculated amplitude of the single photon transition, (b) the calculated amplitude of the two photon transition, and (c) the calculated amplitude of the three photon transition versus the probe detuning $\Delta_p/\Gamma$. The parameters used in the calculations are $g\sqrt{N}=10\Gamma$, $\kappa=2\Gamma$, and $\alpha_p=0.2$.

$|\Psi_\pm^1\rangle=\frac{1}{\sqrt{2}}(|1_-\rangle\pm|s,0\rangle)$ and the coupling laser 2 creates two higher-order dressed polariton states $|\Psi_\pm^2\rangle=\frac{1}{\sqrt{2}}(|2_-\rangle\pm|s,1\rangle)$, which results in the destructive interference for the single-photon excitation and the two-photon excitation as shown in Fig. 3(c). But the three-photon excitation

$|-\frac{N}{2},0\rangle$ - $|3_-\rangle$ is intact and resonantly enhanced, which shows up as a peak in the probe excitation spectrum at $\Delta_p = -g\sqrt{N}$ as shown in Fig. 5(c).

Fig. 4 plots separately (a) the amplitude of the single photon excitation, (b) the amplitude of the two photon excitation, and (c) the amplitude of the three photon excitation by the probe laser versus the probe frequency detuning $\Delta_p/\Gamma$. It shows that at $\Delta_p = -g\sqrt{N}$, both the single photon excitation and the two photon excitation are suppressed, but the three-photon excitation is resonantly enhanced. The other spectral peaks at $\Delta_p = -g\sqrt{N} \pm \Omega_1$ (representing the excitation of the dressed polariton states $|\Psi_\pm^1\rangle = \frac{1}{\sqrt{2}}(|1_-\rangle \pm |s,0\rangle)$), $\Delta_p = -2g\sqrt{N}$ (representing the two-photon Raman transition $|-\frac{N}{2},0\rangle$ - $|1-\rangle$ - $|s,0\rangle$ with a single photon from the probe and a single photon from the coupling 2, which is detuned from the intermidiate polariton state $|1-\rangle$), and $\Delta_p = g\sqrt{N}$ (representing the single-photon excitation to the polariton state $|1+\rangle$) are all dominated by the single-photon transitions.

III. Experimental results

The experiment is done with cold $^{85}$Rb atoms confined in a magneto-optical trap (MOT) produced at the center of a 10-ports stainless-steel vacuum chamber [33]. A tapered-amplifier diode laser (TA-100, Toptica) with output power ~300 mW is used as the cooling laser and supplies three perpendicular retro-reflected beams. An extended-cavity diode laser with an output power of ~20 mW is used as the repump laser. The trapped $^{85}$Rb atom cloud is ~ 1.5 mm in diameter. The three-level atomic system is realized with the Rb D1 transitions in which the ground hyperfine states F=2 and F=3 are chosen as the state $|g\rangle$ and $|s\rangle$, respectively, and the excited hyperfine states F'=3 is chosen as the excited states $|e\rangle$. The decay rate of the excited state $|e\rangle$ is $\gamma \approx 6$ MHz. The standing-wave cavity consists of two mirrors of 5 cm curvature with a mirror separation of ~ 5 cm and is mounted on a stainless holder enclosed in the vacuum chamber. The empty cavity finesse is measured to be ≈500 (the decay linewidth is $\kappa \approx$ 6 MHz). Movable anti-Helmholtz coils is used so the MOT position can be finely adjusted to coincide with the cavity center. Three extended-cavity diode lasers operating at 795 nm are used as the probe laser (couples the F=2-F'=3 transitions) and the two coupling lasers (drives the F=3-F'=3 transitions). The coupling lasers are $\sigma_+$ polarized (the quantization axis is defined as the propagation direction of the coupling lasers, which is perpendicular to the cavity axis) and have a beam diameter of ~ 5 mm, and are made to co-propagate perpendicularly to the intra-cavity probe beam to intercept the cold Rb atoms at the cavity center. The attenuated probe beam is $\pi$ polarized and is coupled into the cavity through a mode-matching lens. The cavity-transmitted probe light passes through an iris and is coupled into a multi-mode fiber, the output of which is collected by a photon counter (PerkinElmer SPCM-AQR-16-FC).

The experiment was run sequentially with a repetition rate of 10 Hz. All lasers were turned on or off by acousto-optic modulators (AOM) according to the time sequence described below. For each period of 100 ms, ~98.9 ms was used for cooling and trapping of the $^{85}$Rb atoms, during which the trapping laser and the repump laser were turned on by two AOMs while the coupling lasers and the probe laser were off. The time for the data collection lasted ~ 1.1 ms, during which the repump laser was turned off first, and then after a delay of ~0.1 ms, the trapping laser was turned off (the current to the anti-Helmholtz coils of the MOT was always kept on), and the coupling lasers and the probe laser were turned on. The probe laser frequency was then scanned across the $^{85}$Rb D$_1$ F=2 to F'=3 transitions and the probe light transmitted through the cavity was recorded versus the probe frequency detuning.

The measured probe light intensity transmitted through the cavity versus $\Delta_p$ is plotted in Fig. 5. Fig. 5(a) shows the probe excitation spectrum without the free-space coupling fields (both $\Omega_1=0$ and $\Omega_2=0$). It exhibits two peaks located at $\Delta_p = \pm g\sqrt{N}$,

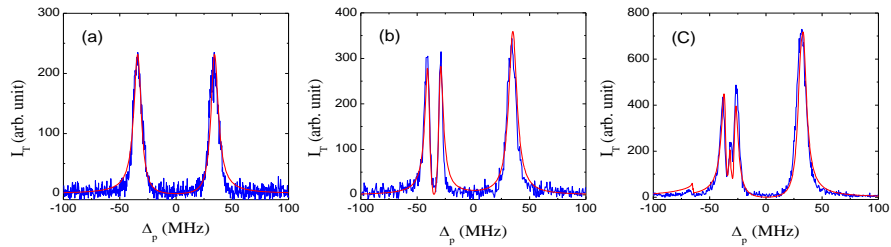

Fig. 5 (a) The probe light intensity transmitted through the cavity versus the probe frequency detuning $\Delta$p. Blue lines are experimental data and red lines are calculations. (a) Without the two coupling fields ($\Omega_1=0$ and $\Omega_2=0$). (b) With only coupling field 1 ($\Omega_1 \approx 10$ MHz and $\Omega_2=0$). (c) With both coupling fields ($\Omega_1 \approx \Omega_2 \approx 10$ MHz), $\Delta_1 = -g\sqrt{N}$ and $\Delta_2 = -2g\sqrt{N}$. The parameters used for the calculations are $\Delta_c = 0$, $g\sqrt{N} = 36$ MHz, $\Delta_1 = -g\sqrt{N}$ and $\Delta_2 = -2g\sqrt{N}$, $\kappa \approx \Gamma = 6$ MHz.

representing the single-photon excitation of the two polariton states $|1-\rangle$ and $|1+\rangle$ (the normal modes) separated in frequency by the vacuum Rabi splitting $2g\sqrt{N}$. Fig. 5(b) shows that when there is only the coupling field 1 with $\Delta_1 = -g\sqrt{N}$, the probe excitation at $\Delta_p = -g\sqrt{N}$ is suppressed by the destructive interference induced by the coupling laser 1. Fig. 5(c) plots the probe excitation spectrum when both coupling fields are present and the detunings are $\Delta_1 = -g\sqrt{N}$ and $\Delta_2 = -2g\sqrt{N}$. The dip at $\Delta_p = -g\sqrt{N}$ in Fig. 4(b) is now turned into a peak, representing the three-photon excitation of the 3rd-order quantum state $|3,-\rangle$ of the CQED system. All other peaks, including the two peaks at $\Delta_p = -g\sqrt{N} \pm \Omega_1$ (the excitation of the first-order dressed state $|\Psi_\pm^1\rangle = \frac{1}{\sqrt{2}}(|1_-\rangle \pm |s,0\rangle)$), a small peak at $\Delta_p = -2g\sqrt{N}$ (the Raman peak), and a peak at $\Delta_p = g\sqrt{N}$ (represent the first-order polariton state $|1+\rangle$) are all dominated by the single-(probe) photon excitations (see Fig. 4 and discussions there).

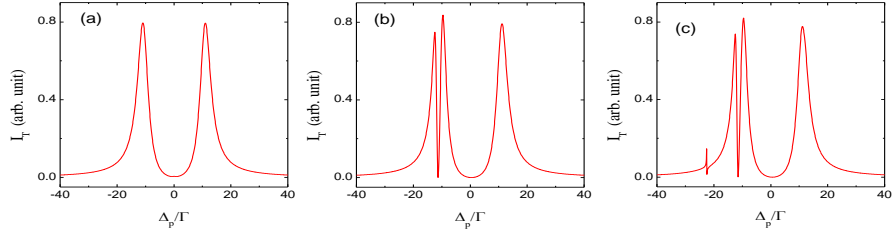

Fig. 6 (a) Cavity-transmitted probe intensity versus the probe frequency detuning $\Delta p$ calculated from the semiclassical analysis. The used parameters are $\Delta_c = 0$, $g\sqrt{N} = 11\Gamma$, $\Delta_1 = -g\sqrt{N}$ and $\Delta_2 = -2g\sqrt{N}$ $\kappa = \Gamma = 6$ MHz. (a) Without the two coupling fields ($\Omega_1 = 0$ and $\Omega_2 = 0$). (b) With only the coupling field 1 ($\Omega_1 = 2\Gamma$ and $\Omega_2 = 0$). (c) With both coupling fields ($\Omega_1 = \Omega_2 = 2\Gamma$), $\Delta_1 = -g\sqrt{N}$ and $\Delta_2 = -2g\sqrt{N}$.

In order to confirm that the observed three-photon peak at $\Delta_p = -g\sqrt{N}$ shown in Fig. 5(c) is a pure quantum phenomenon, we carried out semiclassical CQED calculations, in which the free-space coupling fields, the probe field, and the cavity field are all treated classically. The results are plotted in Fig. 6. The semiclassical calculations agree with the quantized analysis of Fig. 5 for all observed spectral peaks except the one and only pure three-photon peak at $\Delta_p = -g\sqrt{N}$ in Fig. 5(c). The fact that the semiclassical calculations presented in Fig. 6(c) fails to reproduce the spectral peak at $\Delta_p = -g\sqrt{N}$ in Fig. 5(c) confirms that the small spectral peak observed at $\Delta_p = -g\sqrt{N}$ in Fig. 5(c) is solely from the three-photon excitation and represents the observation of the pure quantum feature of the CQED system.

IV. Frequency dependence of the nonlinear cavity QED

When $\Delta_1 = -g\sqrt{N}$ and $\Delta_2 = -2g\sqrt{N}$, the two free-space coupling fields induce the quantum interference that suppresses both the single-photon and two-photon excitations, but leaves the three-photon excitation resonantly enhanced. Here we show that although the nonlinear cavity QED phenomenon is enabled by the coupling-induced interference, it is not associated with cavity electromagnetically induced transparency (EIT) reported in earlier studies [34-37]. The cavity EIT occurs when a free-space coupling laser is tuned to be resonant with the transition of the bare atomic states |s> - |e>, which results in a narrow transmission peak of the probe laser at $\Delta_p$=0. For the CQED system coupled by two free-pace coupling fields, cavity EIT will be created when either one of the two coupling fields is tuned to the atomic resonance,. We have performed the experiments and the theoretical calculations by setting the frequency detunings of the two coupling lasers to $\Delta_1 = -g\sqrt{N}$ and $\Delta_2 = 0$, the results are presented in Fig. 7

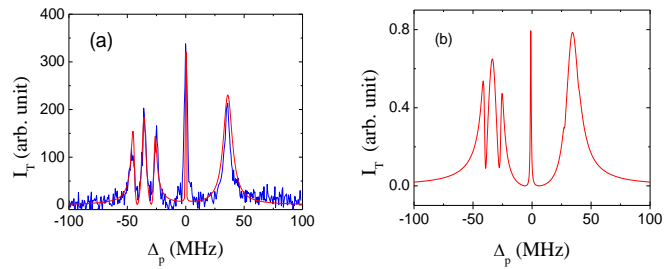

Fig. 7 (a) Cavity-transmitted probe intensity versus the probe frequency detuning Δp. Blue lines are experimental data and red lines are calculations from the quantized analysis with the truncated states. (b) Calculated transmission intensity of the probe laser from a semiclassical analysis. The parameters used in the calculations are

$\Omega_1 \approx \Omega_2 \approx 10$ MHz, $g\sqrt{N} = 36$ MHz, $\Delta_1 = -g\sqrt{N}$, $\Delta_2 = 0$, $\Delta_c = 0$, $\alpha_p$=0.2, and $\kappa \approx \Gamma = 6$ MHz.

Fig. 7(a) plots the measured transmission spectrum of the probe laser under the cavity EIT condition. We observe the signature cavity EIT peak at $\Delta_p = 0$, a spectral peak at $\Delta_p = g\sqrt{N}$ (corresponding to the single-photon excitation of the polariton state $|0,0\rangle - |1+\rangle$), two spectral peaks at $\Delta_p = -g\sqrt{N} \pm \Omega_1$ (corresponding to the single-photon excitation of the dressed polariton states $|\Psi_\pm^1\rangle = \frac{1}{\sqrt{2}}(|1_-\rangle \pm |s,0\rangle)$), and finally a spectral peak at $\Delta_p = -g\sqrt{N}$ that is mostly excited by the single-photon process as shown in Fig. 8 below. This peak is to be distinguished from the pure three-photon excitation peak in Fig. 5(c) even though the peak occurs at exactly the same probe frequency $\Delta_p = -g\sqrt{N}$.

In order to confirm that the observed peak at $\Delta_p = -g\sqrt{N}$ in Fig. 7(a) is mainly from the single-photon process, we performed the semiclassical CQED calculation under the identical conditions in which the free-space coupling fields, the probe field, and the cavity field are all treated classically. The semiclassical calculation is plotted in Fig. 7(b) qualitatively reproduces all observed spectral peaks and thus confirms the single-photon nature of the spectral peaks presented in Fig. 7(a).

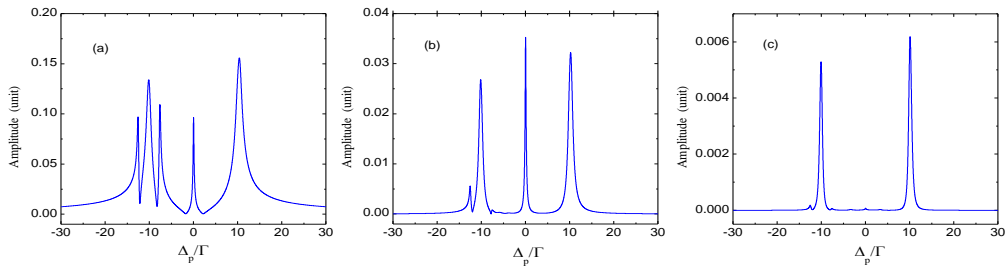

Fig. 8 With $\Delta_1 = -g\sqrt{N}$ and $\Delta_2 = 0$ (cavity EIT is created), (a) the calculated amplitude of the single photon transition versus the probe frequency detuning $\Delta_p$. (b) The calculated amplitude of the two photon transition versus $\Delta_p$. (c) The calculated amplitude of the three photon transition versus $\Delta_p$. It shows that at $\Delta_p = -g\sqrt{N}$, the single photon transition $|-\frac{N}{2},0\rangle - |1_-\rangle$, the two photon transition $|-\frac{N}{2},0\rangle - |1_-\rangle - |2_-\rangle$,

and the three-photon transition $|-\frac{N}{2},0\rangle - |1_-\rangle - |2_-\rangle - |3_-\rangle$ are all resonant, but the single photon transition is dominant. The parameters used in the calculations are the same as that in Fig. 7(a).

As a further confirmation, we also calculated separately the amplitudes of single-photon transition, two-photon transition, and three-photon transition with the quantized analysis (Eq. (7) and Eq. (8)) under the cavity EIT condition $\Delta_1 = -g\sqrt{N}$ and $\Delta_2 = 0$. The results are plotted in Fig. 8. There are spectral peaks at $\Delta_p = -g\sqrt{N}$, $\Delta_p = -g\sqrt{N} \pm \Omega_1$, $\Delta_p = 0$, and $\Delta_p = g\sqrt{N}$. In particular, the peak at $\Delta_p = 0$ is due to the cavity EIT (the excitation of the intra-cavity dark state) [31-34]. However, all of the spectral peaks are dominated by the single (probe) photon excitation. The three-photon excitation amplitude is orders of magnitude smaller and cannot be directly inferred.

V. Conclusion

In conclusion, we have proposed a method to study the quantum nonlinear CQED. The method uses the quantum interference induced by two free-space laser fields to suppress the single-photon excitation and two-photon excitation in the CQED system and resonantly enhances the three photon excitation of the $3^{rd}$-order quantum ladder state. We observed the interference controlled multi-photon excitation of the CQED system in an experiment performed with cold Rb atoms confined in an optical cavity and the experimental results agree with the calculations from a fully quantized analysis based on the truncated state basis. The semiclassical analysis performed for the CQED system cannot reproduce the spectral peak associated with the pure three-photon excitation process, but agree with the experimental measurements and the quantized analysis for the spectral peaks associated with the excitation processes involving only a single probe photon. Thus this represents a direct observation of a pure quantum phenomenon in the multiatom CQED system. It will be interesting to quantify the quantum statistical behavior of the nonlinear excitation process and explore its possible application for the nonclassical light generation and exotic quantum-state preparation.

Acknowledgement

This work is supported by the National Science Foundation under Grant No. 1205565. G. Li acknowledges financial support from the National Basic Research Program of China (Grant No. 2012CB921602). G. Yang acknowledges support from the National Natural Science Foundation of China under Grant No. 11104243.

photon), we refer to the multiphoton transitions here as those involving more than one cavity photon: i.e., the two-photon transition involves two cavity photons, three photon transition involves three cavity photons, and etc.